\documentclass[prl,twocolumn,showpacs,groupaddress]{revtex4}
\usepackage{graphicx,amsfonts,amsmath,amssymb}

\usepackage{epstopdf}
\usepackage{times}
\usepackage{txfonts}

\begin{document}
\bibliographystyle{h-physrev3}
\title{Bright Matter-Wave Soliton Collisions in a Harmonic Trap: Regular and Chaotic Dynamics}
\author{A. D. Martin}
\affiliation{Department of Physics, Durham University, Durham DH1 3LE, United Kingdom}
\author{C. S. Adams}
\affiliation{Department of Physics, Durham University, Durham DH1 3LE, United Kingdom}
\author{S. A. Gardiner}
\affiliation{Department of Physics, Durham University, Durham DH1 3LE, United Kingdom}

\date{\today}
\begin{abstract}
Collisions between bright solitary waves in the 1D Gross-Pitaevskii equation with a harmonic potential, which models a trapped atomic Bose-Einstein condensate, are investigated theoretically. A particle analogy for the solitary waves is formulated, which is shown to be integrable for a two-particle system. The extension to three particles is shown to support chaotic regimes. Good agreement is found between the particle model and simulations of the full wave dynamics, suggesting that the dynamics can be described in terms of solitons both in regular and chaotic regimes, thus presenting a paradigm for chaos in wave-mechanics. 
\end{abstract}

 \maketitle 
The presence of chaos in quantum systems is a topic of intense theoretical and experimental interest \cite{Gutzwiller_Book_1990}. A signature of classical chaos is the ergodic filling of regions in phase space. Applying this criterion in the search for chaos in wave-mechanical systems, for example the linear Schr\"odinger equation in quantum mechanics, one confronts the uncertainty relations, which dictate that trajectories are smeared out. In particular, chaos is impossible to observe when dispersion dominates over the exponential divergence of neighboring trajectories.
For this reason, non-dispersive waves such as solitary waves or solitons
are of particular interest in the study of chaotic dynamics.
In this case, particle-like chaotic behaviour may be well-defined in wave-mechanical systems.

Solitary waves may be found in solutions to nonlinear wave equations where the nonlinearity counteracts the dispersion of a wave-packet such that it retains its form as it propagates.  
The nonlinear Schr\"odinger equation (NLSE) is such an equation, employed to describe diverse physical systems; notably light propagating in fibre-optics \cite{Haus_RMP_1996}, and more recently as an approximation to the many-body dynamics of dilute atomic Bose-Einstein condensates (BEC) \cite{Dalfovo_RMP_1999}, where it is called the Gross-Pitaevskii equation (GPE). 
Solitons are solitary waves that emerge unscathed from collisions, up to shifts in position and phase \cite{Zakharov_JETP_1972}; this behaviour is reminiscent of particle behaviour, motivating the particle-like name soliton. 
The homogeneous 1D NLSE with attractive nonlinearity supports bright soliton solutions \cite{Zakharov_JETP_1972}, so-called because they represent a peak (rather than a trough) in intensity in a nonlinear-optical setting, or  in particle density in BEC.
Macroscopic quantum states consisting of multiple bright matter-wave solitary waves present an interesting testing ground for wave chaos. In general all wave-packet evolution predicted by the Schr\"odinger equation is periodic or quasiperiodic, due to its linearity. The nonlinearity in the 1D GPE and associated solitary wave solutions may break all periodicity, leading to the realization of ergodic behaviour, i.e., recognizably chaotic dynamics.

Bright solitary waves have been the subject of substantial experimental and theoretical investigation in the context of nonlinear optics \cite{Haus_RMP_1996,Scharf_PRA_1992,Scharf_CSF_1995}, and BEC
\cite{Carr_PRL_2004,Strecker_Nature_2002,Khaykovich_Science_2002,Cornish_unpublished_2006,Parker_unpublished_2006}. Notably, chaotic and regular soliton behaviour have been observed theoretically in a NLSE with a $\delta$-kicked rotor potential \cite{Benvenuto_PRA_1991}. In BEC experiments, the magnetic or optical trap employed to confine the constituent atoms introduces a position-dependent potential. In this Letter, we investigate to what extent solitary wave collisions in a harmonic potential provide a paradigm for particle-like chaotic behaviour in a wave-mechanical system. Due to this potential, bright solitary waves in a harmonically trapped system are not true solitons; however, it will be shown that the particle nature of the solitary waves is very pronounced, so the bright solitary waves in this system will from now on be called solitons. In order to test the extent of the soliton behaviour, we introduce a particle model, adapted from a nonlinear optics context for a NLSE with a sinusoidal external potential \cite{Scharf_PRA_1992}. In this model, constructed for the regime where the solitary waves are well-separated before and after collisions, the waves are modeled as interacting classical particles. Within this model, we show that the two soliton case is integrable, but for three (or more) solitons one can expect chaotic dynamics. The results are compared to numerical solutions of the NLSE, and provide a probe of the coexisting particle and wave properties of bright solitary waves. The most surprising result is that the particle-like behaviour is preserved even in the chaotic regime. In contrast to the linear Schr\"odinger equation, where the evolution of localized wave-packets is rapidly disrupted in regimes supporting classically chaotic dynamics \cite{Gutzwiller_Book_1990}, the soliton solutions appear to be remarkably robust. 
\begin{figure}[tbp]
\centering
\includegraphics[width=8.8cm]{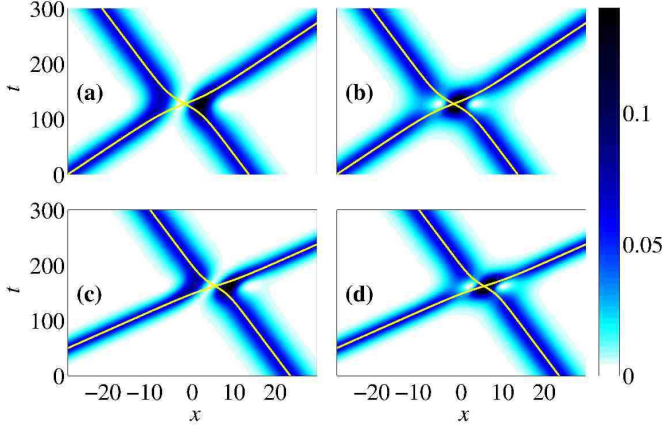}
\caption{(Colour online). Trajectories in the particle model (yellow lines) plotted over density distributions predicted by 1D dynamics in the homogeneous GPE. The trajectories correspond to solitons colliding with a relative phase of the golden ratio $\phi=(1+\sqrt{5})/2$ in (a) and (c), and a relative phase of $\phi=\pi(1+\sqrt{5})/2$ in (b) and (d). In (a) and (b) the incoming speeds of the solitons are -0.1 $|g_{1D}|N/\hbar$ and 0.2 $|g_{1D}|N/\hbar$, and in (c) and (d) the incoming speeds are -0.1 $|g_{1D}|N/\hbar$  and 0.3 $|g_{1D}|N/\hbar$. Taking the parameters of the system to agree with recent experiment \cite{Strecker_Nature_2002}, the solitons have equal effective masses, the axial trapping frequency is 10 Hz, and other parameters (radial trap frequency of 800 Hz, atomic mass and scattering length of $^{7}$Li, and 5000 particles per soliton). The unit of $x$ is then equal to 3.6 $\mu$m, and a unit of $t$ to 1.4 ms.} \label{fig_collision}
\end{figure}
At temperatures encountered in atomic BEC experiments (nK or lower), the atom-atom interaction potential may generally be replaced by an effective contact interaction, quantified by the $s$-wave scattering length, $a$. Depending on species, this may be positive or negative. By exploiting a Feshbach resonance, it may also be tuned using an external magnetic field \cite{Inouye_Nature_1998}. In the case of a trapped, almost fully Bose-condensed dilute atomic gas, the dynamics are largely governed by the following GPE:
\begin{equation}
i \hbar \frac{\partial}{\partial t} \Psi(\mathbf{r},t)=
\left[
-\frac{\hbar^{2}\nabla^{2}}{2m}+V_{\mbox{\scriptsize
ext}}(\mathbf{r})+g_{3D}N\left|\Psi(\mathbf{r},t) \right|^{2}
\right]
\Psi(\mathbf{r},t),
\label{GP}
\end{equation}
where $N$ is the total number of atoms, $m$ the atomic mass, and $\Psi(\mathbf{r},t)$ the condensate mode-function, normalised to one.  The atom-atom interactions are quantified by $g_{3D}=4\pi\hbar^{2}a /m$, where, in this Letter, $a$ is negative. The proportion of non-condensate atoms is thus assumed to be negligible; it can be shown, however, that linear instabilities in the GPE directly imply \cite{Bogoliubov_JPhys_1947,Castin_PRA_1998} that the population of the non-condensate component may rapidly become significant.  Regimes where soliton collision dynamics have a chaotic character are thus of additional interest, as they may coincide with a greater tendency for linear instability, and hence implicitly with condensate depletion \cite{Gardiner_JMO_2002}.
\begin{figure}[tbp]
\centering
\includegraphics[width=6.0cm,angle=270]{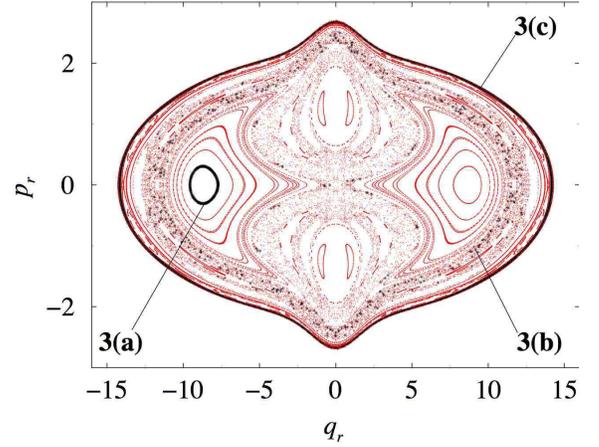}
\caption{(Colour online).  Poincar\'e section of the three-soliton system with $\tilde{H}$=10.
Regions corresponding to trajectories in figures 3(a) to 3(c) are labeled, and highlighted using larger, darker points. The section corresponds to the momentum $p_{r}$ and position $q_{r}$ of the ``asymmetric stretch'' mode when the ``stretch'' mode coordinates $q_{c}=0$, $p_{c}<0$.
The figures correspond to the regime where the solitons have equal effective masses, the axial trapping frequency is 10 Hz, and the other parameters (radial trap frequency of 800 Hz, atomic species mass and scattering length of $^{7}$Li, and 5000 particles per soliton) correspond to recent experiment \cite{Strecker_Nature_2002}.} \label{fig_poin}
\end{figure}

The regime of interest for the study of solitons is the quasi-1D case, where the atoms are trapped in a radially tight harmonic trap with loose harmonic axial confinement.  We may assume a harmonic ground state (Gaussian) ansatz in the radial direction, since the harmonic potential energy dominates over the interaction energy. The GPE then takes the form:
\begin{equation}
i\frac{\partial}{\partial t}\psi(x)=-\frac{1}{2}\frac{\partial^{2}}{\partial x^{2}}\psi(x)+\frac{\omega^{2}x^{2}}{2}\psi(x)-
|\psi(x)|^{2}\psi(x) \label{S2},
\end{equation}
where $x$ is now measured in units of
$\hbar^{2}/m|g_{1D}|N$ and
$t$ in units of
$\hbar^{3}/m|g_{1D}|^{2}N^{2}$
with $g_{1D}=2\hbar \omega_{r}a$, $\omega_{r}$ is the radial trapping frequency and $\omega$ is the axial frequency in our units of inverse time.
In the case of vanishing harmonic potential, an exact solution exists \cite{Zakharov_JETP_1972}, comprising an arbitrary number of well separated solitons taking the form:
\begin{equation} \Phi_{i}(x,t)=2\eta_{i} \mathrm{sech} \left[2\eta_{i}(x-q_{i})\right]
e^{i v_{i}(x-q_{i})}e^{i(2\eta_{i}^{2}+v_{i}^{2}/2) t}e^{i\alpha_{0i}},
\label{Sol_1}
\end{equation}
where $q_{i}=v_{i}t +x_{0i}$ is the position of the peak of the $i$th soliton; $x_{0i}$ is the peak position at $t=0$; $\alpha_{0i}-v_{i}x_{0i}$ is the phase at $x=0$, $t=0$; and $v_{i}$ are the soliton velocities. Our normalisation condition implies that
$\sum_{i}^{N_{s}}4\eta_{i}=1$, where $N_{s}$ is the number of solitons present.
When these solitons emerge from collisions, they suffer position shifts dependent on their initial speeds $v_{i}$, and the quantities $\eta_{i}$ known as effective masses only. The outgoing soliton motion does not depend on the relative phase \cite{Zakharov_JETP_1972}. This property is illustrated in Fig. 1 where we compare the solution of the 1D GPE with a particle model which is outlined below. Note that the density distribution during the collision does depend on the relative phase, but this does not influence the outgoing trajectories.

The particle model follows the approach of Scharf and Bishop \cite{Scharf_PRA_1992}, which reproduces the position shifts following collisions, and also reproduces the motion due to the trapping potential, whilst neglecting the phase behaviour. This approach is appropriate for regimes where the solitons are well separated between collisions. This approach is not appropriate for soliton trains, as observed by Strecker \textit{et al.} \cite{Strecker_Nature_2002}, and modeled by Gordon \cite{Gordon_OptLett_1983} in the absence of any external potential, and by Gerdjikov \textit{et al.} \cite{Gerdikov_unpublished_2005} in the case of harmonic confinement, where the solitons are never well separated, and the phase difference has an important effect. Parker \textit{et al.} have modelled bright matter-wave  soliton collisions \cite{Parker_unpublished_2006} using the 3D GPE at or near the quasi-1D regime. Taking the full 3D GPE dynamics into account highlights some important deviations from the 1D dynamics: in particular, collapse may occur during collisions of solitons having slow approach speeds, with sensitivity to the relative phase of the solitons. However, above a particular threshold velocity, the quasi-1D model considered in this Letter can be expected to hold, including the observed phase-insenstivity of the soliton collision dynamics.

Following reference \cite{Scharf_PRA_1992}, the action of the external potential is deduced by using the one soliton solution of the homogeneous case as an ansatz for the system with a harmonic potential, and evaluating the norm and energy functionals, which are known to be constant. This gives equations of motion for the solitons in the external potential. Added to this is an inter-particle potential which reproduces the position shifts of the solitons on emerging from collisions with each other, inferred from the exact solution to the homogeneous NLSE, which are assumed not to change upon the addition of the position-dependent external potential. The Hamiltonian for an arbitrary number of solitons
($N_{s}$) is given in the particle analogy by:
\begin{equation}
\begin{split}
H= & \sum^{N_{s}}_{i=1}\left(\frac{p_{i}^{2}}{2\eta_{i}} + \frac{\eta_{i} \omega^{2} q^{2}_{i}}{2}\right)
\\&-\sum_{1\leq i< j\leq
N_{s}} 2\eta_{i}\eta_{j}(\eta_{i}+\eta_{j})\mathrm{sech}^{2}\left[\frac{2\eta_{i}\eta_{j}}{\eta_{i}+\eta_{j}}(q_{i}-q_{j}) \right],  \label{Ham4}
\end{split}
\end{equation}
in which the reason for the interpretation of $\eta_{i}$ as effective masses is clear.
This Hamiltonian models the positional dynamics of the soliton peaks. In the case of two solitons ($N_{s}=2$) with identical effective masses, the following coordinates may be defined: the centre-of-mass position $ Q:=(q_{1}+q_{2})/2 $ and the relative
position $q:=q_{1}-q_{2}$. The Hamiltonian [Eq. \ref{Ham4}] reduces to 
\begin{equation}
\begin{split}
 H=&\frac{P^{2}}{4
\eta} +\eta \omega^{2}Q^{2} +\frac{p^{2}}{\eta}+\frac{ \eta\omega^{2}q^{2}}{4}-4\eta^{3}\mathrm{sech}^{2}\left(\eta q\right),
\end{split}
\end{equation} where $P$ is the momentum conjugate to $Q$, and $p$ the momentum
conjugate to $q$. The Hamiltonian is seperable into two parts: the centre-of-mass energy $E$ (dependent on $P$ and $Q$ only), and the ``relative energy'' $\epsilon$ (dependent on $p$ and $q$ only). There are thus two independent constants of the motion, $E$ and $\epsilon$, as many as there are degrees of freedom. Hence, the particle model for two solitons is integrable and the dynamics must be completely regular. The same argument holds in the simple generalisation to non-identical masses.

In the case of three solitons ($N_{s}=3$), the situation is different. When the masses are identical, a useful coordinate change may be made to $Q_{T}/\eta:=(q_{1}+q_{2}+q_{3})/3$, the centre-of-mass position, and $q_{c}/\eta:=(q_{1}-q_{3})/2$  (corresponding to the ``stretch'' mode) and
$q_{r}/\eta:=(q_{1}+q_{3}-2q_{2})$ (corresponding to the ``asymmetric stretch'' mode), the normal coordinates of the system for small displacements from the origin. The stretch modes are similar to the vibrational modes in a tri-atomic molecule \cite{Tennyson_ChemPhys_1985}; as the system is constrained to 1D, however, there is no analogue of the molecular bending mode. Rescaling time
$\tilde{t}=\eta^{2}t$ , and introducing the momenta $p_{c}=2\dot{q_{c}}$ and $p_{r}=\dot{q_{r}}/6$, we may remove the centre-of-mass behaviour from the
problem, as it decouples from the other degrees of freedom. (An equivalent treatment is possible for non-identical masses, $\eta_{i}$.) The resultant Hamiltonian we call the reduced system Hamiltonian
\begin{equation}
\begin{split}
\tilde{H}= & 3p_{r}^{2}+\frac{\omega^{2}}{2\eta^{4}}\frac{q_{r}^{2}}{12}+\frac{p_{c}^{2}}{4}+\frac{\omega^{2}}{2\eta^{4}}q_{c}^{2}-4\mathrm{sech}^{2}(2q_{c})
\\ &
-4\mathrm{sech}^{2}(q_{c}+\frac{q_{r}}{2})-4\mathrm{sech}^{2}(q_{c}-\frac{q_{r}}{2}). \label{Ham_red}
\end{split}
\end{equation}
This Hamiltonian, describing the two remaining degrees of freedom, is not separable, and it is necessary to integrate the equations of motion numerically to analyse the system's behaviour. Poincar\'e sections illustrate regions of regular and chaotic behaviour. 
Figure \ref{fig_poin} shows a Poincar\'e section corresponding to the momentum $p_{r}$ and position $q_{r}$ of the ``asymmetric stretch'' mode when the ``stretch'' mode coordinates $q_{c}=0$, $p_{c}<0$; other sections would be equally illustrative of the qualitative behaviour. The behaviour is regular at large positive values of $\tilde{H}$, but as $\tilde{H}$ is reduced, chaotic behaviour emerges, characterised by ergodic regions in between regular tori. For small $\tilde{H}$ the system is mostly an ergodic sea, with islands of stability. A regime with both ergodic regions and regular tori is plotted in Fig. \ref{fig_poin}. Consideration of the form of the reduced-system
Hamiltonian [Eq. (\ref{Ham_red})] shows that without the interaction the system is integrable, as it becomes a decoupled pair of harmonic oscillators. When $\tilde{H}$ is large and positive, the interaction part of the Hamiltonian (which is always negative) should give a relatively small contribution to the Hamiltonian, compared to the integrable part of the Hamiltonian (which is always positive). When $\tilde{H}$ is reduced, this is no longer the case, and chaos is emergent.
\begin{figure}[tbp]
\centering
\includegraphics[width=8.8cm]{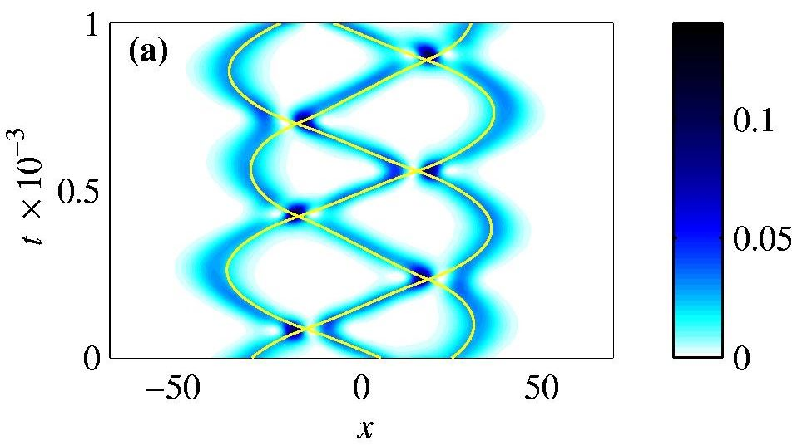}
\includegraphics[width=8.8cm]{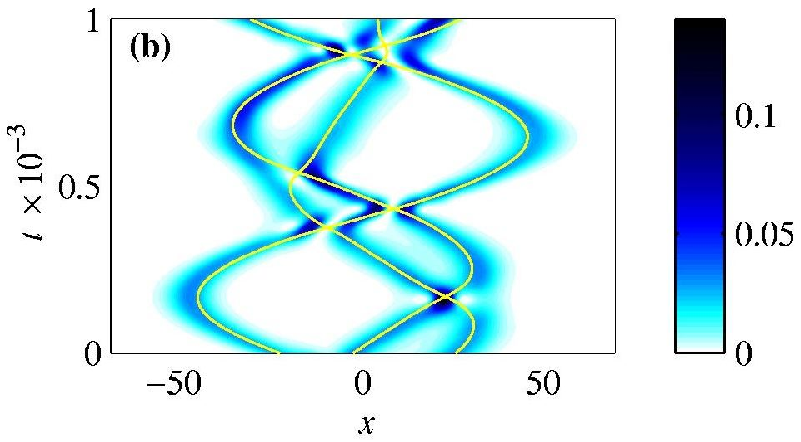}
\includegraphics[width=8.8cm]{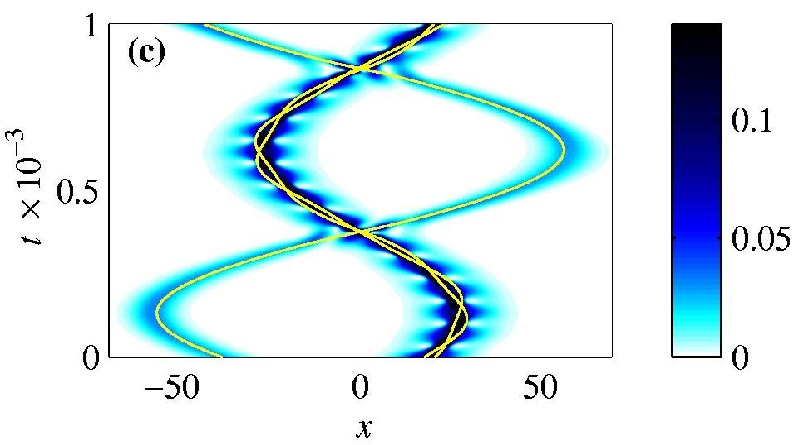}
\caption{(colour online). Trajectories in the particle model (yellow lines) plotted over density distributions predicted by 1D GPE dynamics, corresponding to (a) a regular orbit, (b) a chaotic orbit, and (c) a bound state, as shown in Fig.\ \ref{fig_poin}(b). In the centre of the bound state, the density increases to approximately 0.5 normalised units, but our scale is pinned at 0.14 units in order to resolve the low density regions better. The parameters of the system are $\tilde{H}$=10, the solitons have equal effective masses, the axial trapping frequency is 10 Hz, and other parameters (radial trap frequency of 800 Hz, atomic mass and scattering length of $^{7}$Li, and 5000 particles per soliton) correspond to the recent experiment \cite{Strecker_Nature_2002}. The unit of $x$ is then equal to 2.4 $\mu$m, and a unit of $t$ to 0.6 ms. } \label{fig_traj}
\end{figure}

Figure \ref{fig_traj} shows a comparison of trajectories in the particle model with results from integrations of the 1D GPE [Eq. (\ref{S2})]. The comparisons illustrate the good agreement between the particle model and the 1D GPE in the regimes in which the particle model is valid, i.e., when solitons are well separated between collisions [Figs. \ref{fig_traj}(a) and \ref{fig_traj}(b)], even when the motion is chaotic [Fig. \ref{fig_traj}(b)]. When two of the solitons are not well separated [Fig. \ref{fig_traj}(c)], the 1D GPE simulation shows that a ``bound state'' is formed, which looks like a single ``higher-order'' soliton with an excited breathing mode. The particle model does not predict well the behaviour within the ``bound state'', but does give a good prediction of the centre-of-mass motion of the ``bound state'' and its interactions with the other soliton; it is likely that the behaviour of the density of the ``bound state'' is strongly coupled to the phase behaviour within the ``bound state''. 

These results show that harmonically trapped solitons have strong particle characteristics. Even in chaotic regimes, where the exponential growth of linear instabilities is most prevalent, the soliton solution is remarkably robust. An echo of the wave-equation origin of the particle model is that, consistent with the attractive interaction potential, the particles pass through each other subsequent to collisions. The payoff of using the particle model is that it is quicker to solve than the 1D GPE (involving four real variables, rather than a continuous complex-valued function); thus Poincar\'e sections (see Fig. \ref{fig_poin}) can rapidly build up a qualitative idea of the many-soliton behaviour. 

The experimental demonstration of such chaotic dynamics in a wave mechanical system requires a relatively straightforward adaption of recent experiments on bright matter wave solitons  \cite{Strecker_Nature_2002,Khaykovich_Science_2002,Cornish_unpublished_2006}. For example, a system of 3 solitons can be created reproducibly by careful choice of the initial conditions \cite{Cornish_unpublished_2006}. Manipulation of the optical trapping potential during the creation of the solitons will allow the solitons' initial velocities to be chosen. The chaotic regions of phase space may be probed by measuring the sensitivity of the subsequent evolution of the density distribution to the initial condition.

In conclusion, an effective classical particle model has been derived for many solitons in the NLSE with a harmonic potential. This applies to a dilute BEC of attractive atoms in the quasi-1D limit of a cylindrically-symmetric cigar-shaped trap. Within this model two-soliton dynamics are fully integrable and regular, but three solitons may display chaotic dynamics when atom-atom interactions are significant. The particle model exhibits good agreement with the 1D GPE in the regime of large separation of the solitons before and after collisions, even when the particle motion is chaotic.  This confirms the surprising robustness of bright matter-wave solitons, as has also been observed experimentally  \cite{Strecker_Nature_2002,Khaykovich_Science_2002,Cornish_unpublished_2006}. There is a good degree of agreement even when ``bound states'' are modeled (states not in a regime of large separation). Chaotic regions may also be a useful predictor of regimes of condensate instability, which can be explored with a fuller treatment of the condensate and non-condensate atoms  \cite{Castin_PRA_1998}. 

We thank  J. Brand, S. L. Cornish, T. S. Monteiro and N. G. Parker for useful discussions, and acknowledge support from the UK EPSRC.

\end{document}